\documentclass[trackchanges,twocolumn]{aastex701}

\usepackage{amsmath}
\usepackage{booktabs}
\usepackage{bm}
\usepackage{romannum}

\begin{document}

\title{Measuring the Moving Lens Effect with Wide-Separation Lensed Quasars}

\author{Yupeng Zhang}
\affiliation{School of Physics and Astronomy, Beijing Normal University, Beijing 100875, China}
\email{202321160024@mail.bnu.edu.cn}

\author{Shuxun Tian}
\affiliation{School of Physics and Astronomy, Beijing Normal University, Beijing 100875, China}
\email{tshuxun@bnu.edu.cn}

\author{Zhengxiang Li}
\email[show]{zxli918@bnu.edu.cn}
\affiliation{School of Physics and Astronomy, Beijing Normal University, Beijing 100875, China}

\begin{abstract}

The transverse motion of a gravitational potential induces frequency shifts in photons propagating through it, giving rise to the moving-lens effect and providing a probe of transverse peculiar velocities. Statistical evidence for this effect has recently been reported using cross-correlations between cosmic microwave background temperature maps and large-scale-structure tracers.
Here, we investigate a complementary spectroscopic approach based on wide-separation lensed quasars, for which different lensed images probe distinct deflection geometries through a moving foreground lens. We focus on SDSS J1004+4112 and assess the differential redshift that could be measured from its broad emission-line spectra. Using an estimator that combines information across the line profiles, we perform Monte Carlo forecasts for C~IV and C~III] under an ELT-like configuration, considering photon noise only.
For an assumed lens transverse velocity of $500\,\mathrm{km\,s^{-1}}$ and the orientation maximizing the signal, the expected differential redshift reaches $|\Delta z_{\rm ML}|\simeq6.7\times10^{-7}$. At an integration time of $40\,\mathrm{hr}$, the joint photon-noise-limited precision is $\sigma_z\simeq2.5\times10^{-7}$. C~IV dominates the constraint, while C~III] provides only a modest additional improvement, and integrations of tens of hours reach the $10^{-7}$ precision regime. These results show that photon statistics alone can reach the characteristic moving-lens signal scale, although practical detectability will depend on controlling astrophysical and instrumental systematics.
\end{abstract}

\section{Introduction}
Peculiar velocities provide a direct tracer of the growth of cosmic structure, but observational access to the velocity field is highly anisotropic. Redshifts, redshift-space distortions, and the kinematic Sunyaev--Zel'dovich effect primarily constrain the line-of-sight component, whereas transverse motions of extragalactic systems are difficult to measure directly \citep{2019PhRvL.123f1301H,2021PhRvD.104h3529H}.
The moving-lens effect induces a frequency shift in photons passing through a gravitational potential that moves transversely to the observer's line of sight \citep{1983Natur.302..315B,1986Natur.324..349G,1993ApJ...415..459P}.
For a fixed observer's line of sight, a transversely moving gravitational potential constitutes a time-varying gravitational field, causing photons passing through it to undergo a change in energy \citep{2004PhRvD..69f3001W}. For the cosmic microwave background, this effect appears as a dipolar temperature perturbation with opposite signs on the two sides of the lens; for a background source with well-defined spectral features, it instead manifests as a directly measurable shift in the wavelength or frequency of the spectral lines.
The signal depends on the lens velocity and the light-deflection geometry, providing a way to probe the transverse motion through two complementary observational approaches.

The first approach uses the dipolar temperature signal imprinted on the cosmic microwave background (CMB) around a moving halo, with its orientation set by the transverse motion \citep{1995ApJ...445L..73T,1996ApJ...463...15T,2010MNRAS.407..201C}.
This moving-cluster signal is related to the Rees--Sciama effect \citep{1968Natur.217..511R,2000ApJ...537..542M}.
Statistical methods combining CMB and galaxy surveys have been developed to probe the signal and reconstruct the large-scale transverse velocity field \citep{2019PhRvL.123f1301H,2021PhRvD.104h3529H}, with related applications to the mean pairwise velocity of galaxy clusters \citep{2019ApJ...873L..23Y}.
Foreground mitigation and the associated noise and systematic uncertainties have also been investigated \citep{2025PhRvD.111d3510B}.
A statistical detection of the moving-lens effect has now been reported through CMB/large-scale-structure (LSS) cross-correlation, using Atacama Cosmology Telescope (ACT) CMB temperature maps and galaxies from the DESI Legacy Imaging Surveys \citep{2026arXiv260518938H}.

The second approach uses direct differential spectroscopic measurements between multiple images of a strongly lensed source.  
\citet{2003ApJ...586..731M} first proposed that different light paths through a transversely moving cluster potential acquire different frequency shifts, producing an image-to-image differential redshift that can be used to probe the cluster transverse velocity.  
\citet{2013ApJ...774...70M} further quantified this signal for the Bullet Cluster and discussed its detectability with high-resolution spectroscopy, including possible applications to strongly lensed quasars.  
More recently, \citet{2026arXiv260938457W} revisited the spectroscopic detection of the moving-lens effect in a merging cluster system, further exploring the observational feasibility of this approach.  
Such measurements probe individual strong-lens systems and complement the statistical transverse-motion signal measured through CMB/LSS cross-correlation.

The dependence on transverse motion and deflection geometry makes the moving-lens effect a complementary probe to perspective-rotation methods \citep{2010ApJ...716L.205H}. 
For two lensed images, the differential frequency shift is proportional to the projection of the lens transverse velocity onto the difference between their deflection angles, so systems combining substantial transverse motion with large lensing deflections are particularly favorable targets.
Wide-separation lensed quasars naturally satisfy these requirements: image separations exceeding $10\arcsec$ are typically produced by cluster-scale gravitational potentials, in contrast to the much smaller separations characteristic of galaxy-scale quasar lenses \citep{2007MNRAS.378..469L}.
The first such system, SDSS J1004+4112, was discovered with a maximum image separation of $14.62\arcsec$ and was subsequently confirmed to be lensed by a galaxy cluster \citep{2003Natur.426..810I,2004ApJ...605...78O}; even larger separations are found in systems such as SDSS J1029+2623, whose images span $22.5\arcsec$ \citep{2006ApJ...653L..97I}.
Galaxy clusters also possess substantial peculiar motions \citep{2001MNRAS.322..901S}, while quasars provide bright, compact background sources with prominent emission features suitable for precise differential spectroscopy \citep{2001AJ....122..549V}.
Wide-separation cluster-lensed quasars therefore combine large deflection-angle differences, appreciable lens transverse velocities, and favorable background spectra, making them promising systems for direct spectroscopic measurements of the moving-lens effect.

In this paper, we assess the photon-noise-limited sensitivity of direct differential spectroscopy with a next-generation 30-meter-class telescope under an adopted ELT-like configuration.
Using the $\mathrm{C \ \Romannum{4}}$ and $\mathrm{C \ \Romannum{3}]}$ emission lines of SDSS J1004+4112, we evaluate a 40 hr benchmark and examine the dependence of the differential-redshift precision on integration time.
These calculations provide an idealized benchmark for assessing how the expected moving-lens signal compares with the statistical precision attainable under photon-noise-only assumptions.

In Section~\ref{sec:Derivation of Moving Lens Effect}, we derive the moving-lens differential redshift and estimate its maximum magnitude in known wide-separation lensed quasars.
Section~\ref{sec:Frequency Shift Measurement and Simulation} describes the estimator and mock spectra, and Section~\ref{sec:Results} presents the photon-noise-limited precision and its integration-time dependence.
We summarize the results and discuss observational limitations in Section~\ref{sec:Summary and Discussions}.

\section{Derivation of the Moving-Lens Effect}
\label{sec:Derivation of Moving Lens Effect}

The moving-lens effect can be understood as an observer-dependent consequence of photon deflection in curved spacetime \citep{2003ApJ...586..731M}, as a gravitomagnetic effect \citep{1999PhRvD..60l4002K}, or as energy exchange in scattering by a moving lens \citep{2004PhRvD..69f3001W}.
In general relativity, the measured photon frequency corresponds to the temporal component of its four-wavevector in the observer's rest frame.
For a stationary lens, light has the same frequency before and after traversing the potential in the lens rest frame, while deflection changes the spatial components of the four-wavevector.
If the lens moves relative to the observer, this spatial change projects onto the temporal component in the observer's frame and produces a frequency shift.
We consider only the bulk motion of the lens potential and neglect its intrinsic time variability.
The resulting shift is \citep{2003ApJ...586..731M}
\begin{equation}
    \frac{\nu_{1} -\nu_{0}}{\nu_{0}} = -\gamma \frac{\bm{v}  \bm{\cdot} \hat{ \boldsymbol{\alpha}}}{c}\label{eq:moving_lens} ,
\end{equation}
where $\nu_0$ and $\nu_1$ are the photon frequencies before and after passage through the potential, $\hat{\boldsymbol{\alpha}}$ is the deflection-angle vector in the lens plane, $\gamma\equiv 1/{\sqrt{1-{\bm{|v|}^2}/{c^2}}}$ is the Lorentz factor, $\bm v$ is the lens bulk velocity, and $c$ is the speed of light.
For the cluster velocities considered here, ${\bm {|v|}}\ll c$.
Since the deflection angle lies in the plane of the sky, only the transverse velocity contributes.

\begin{figure}
     \centering
     \includegraphics[width=1\linewidth]{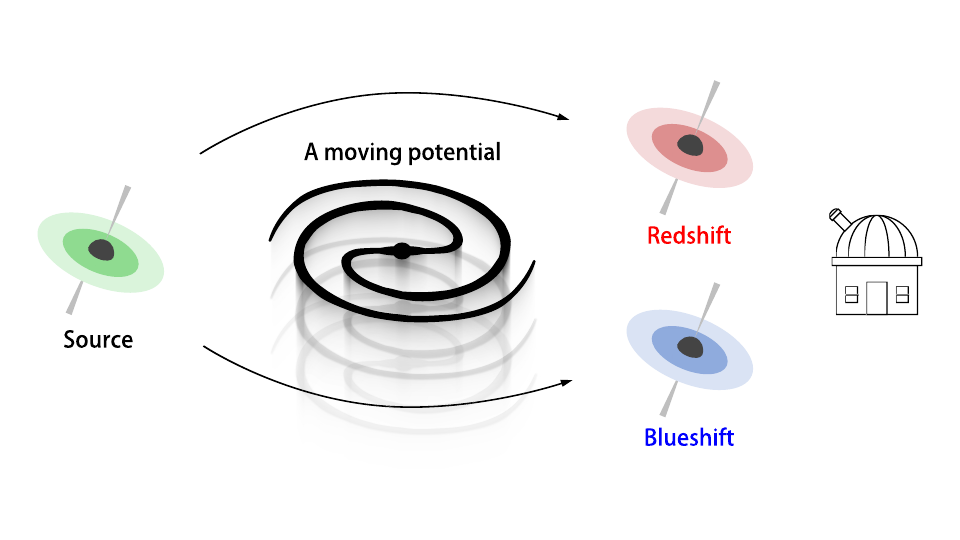}
     \caption{Illustrations for the effect of a gravitational potential with bulk motion.}
     \label{fig:fig1}
 \end{figure}

A transversely moving lens therefore produces different frequency shifts along the paths of its multiple images, as illustrated in Figure~\ref{fig:fig1}.
The images share the source's cosmological redshift but acquire different moving-lens contributions because their deflection angles differ.
Cluster lenses are useful targets because their peculiar velocities can reach several hundred kilometers per second \citep{1996ApJ...462L..49B} and their larger deflections can produce wide image separations.

\subsection{Redshift Difference Estimation}
For a strongly lensed source, the redshift of each image contains the source's cosmological redshift $z_s$ and a contribution $z_m$ from the moving-lens effect.
Subtracting the redshifts of images A and B gives
\begin{align}
    z_A - z_B &= (1+z_s)(z_{Am}-z_{Bm}).
\end{align}
The moving-lens contribution follows from the photon frequency change across the lens.
Using Equation~\eqref{eq:moving_lens}, we obtain
\begin{align}
    z_A - z_B =\frac{(1+z_A)(1+z_B)}{1+z_s}\frac{\gamma}{c}{\bm v} \bm{\cdot} ({ \hat{\boldsymbol{\alpha}}_A} -{ \hat{\boldsymbol{\alpha}}_B} ) ,
\end{align}
Since the moving-lens effect is a second-order contribution \citep{2011Natur.477..567W,2013PhRvD..88d3013Z}, we use $z_s\approx z_A\approx z_B$ on the right-hand side.
We introduce the scaled deflection angle $\boldsymbol{\alpha}=\left({D_{ds}}/{D_s}\right)\hat{\boldsymbol{\alpha}}$, where $D_s$ and $D_{ds}$ are the angular-diameter distances from the observer and the lens to the source, respectively.
The lens equation $\boldsymbol{\beta}=\boldsymbol{\theta}-\boldsymbol{\alpha}$ then gives
\begin{align}
   z_A - z_B &\approx(1+z_s)\frac{\gamma}{c}\frac{D_{s}}{D_{ds}}{\bm v} \bm{\cdot} \boldsymbol{\phi}_{BA} ,
   \label{eq:redshift_in_cosmo}
\end{align}
where $\boldsymbol{\phi}_{BA}$ denotes the angular separation vector between images A and B.
Within this approximation, the difference in deflection angles is determined by the observed image separation, so evaluating the differential redshift does not require a specific lens mass profile.
We denote this moving-lens differential redshift by $\Delta z_{\rm ML}=z_A-z_B$.

Table~\ref{tab:redshift_of_quasars} lists the expected maximum moving-lens differential redshifts for the adopted sample of wide-separation lensed quasars, assuming a cluster transverse velocity of $500\,\mathrm{km\,s^{-1}}$.
The column headed ``Max $\Delta z$'' gives the maximum possible magnitude under this assumption, attained when the transverse velocity is aligned with the image-separation direction and the magnitude of the inner product in Equation~\eqref{eq:redshift_in_cosmo} is largest.
``Max Sep'' denotes the largest angular separation between images in each system.

\begin{figure}[htbp]
    \centering
    \includegraphics[width=0.45\textwidth]{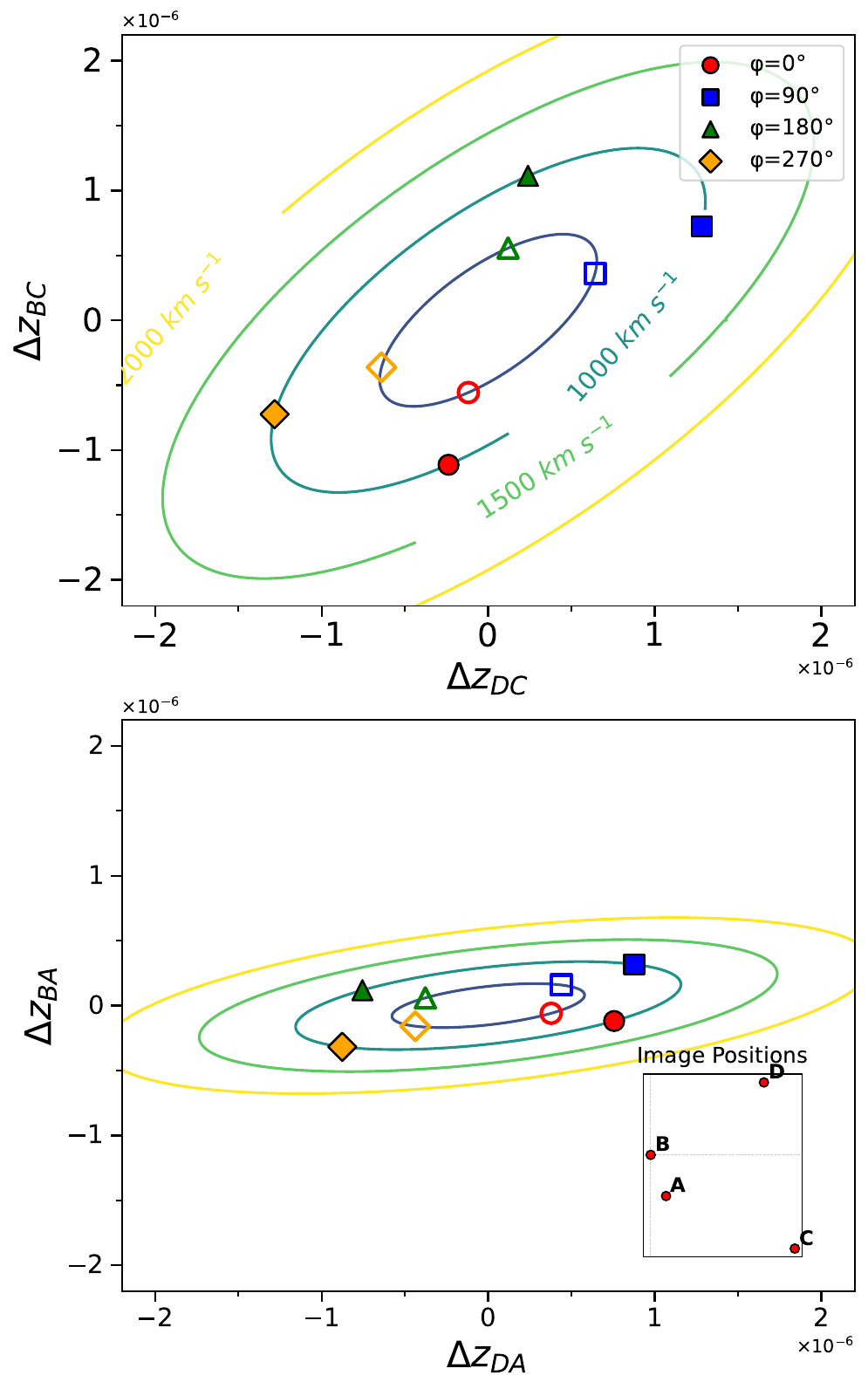}
    \caption{Contour maps of the transverse velocity for SDSS J1004+4112. The axes of the top panel correspond to the redshift difference between image C and image B/D, and the axes of the bottom panel correspond to the redshift difference between image A and image B/D. Different symbol shapes in the legend indicate the angle between the direction of lens motion and the equatorial coordinate system. The inset in the bottom panel shows the relative spatial positions of the different images of SDSS J1004+4112. The maximum separation is between images B and C. The curves correspond to velocities of 500, 1000, 1500, and 2000 $\mathrm{km\ s^{-1}}$, from inner to outer.}
    \label{fig:tran_vel}
\end{figure}

\startlongtable
\begin{deluxetable*}{lcccc}
\tablecaption{The Maximum Redshift Differences Between Images in Different Lens Systems\label{tab:redshift_of_quasars}}

\tablehead{
\colhead{Object} & \colhead{Max Sep [Deg]} & \colhead{Source $z$} & \colhead{Lens $z$} & \colhead{Max $\Delta z$}
}
\startdata
SDSS J1004+4112 \citep{2003Natur.426..810I} & $14\farcs62$ & 1.73 & 0.68 & $0.67\times 10^{-6}$ \\
SDSS J1029+2623 \citep{2006ApJ...653L..97I}& $22\farcs5$& 2.19 & 0.55 & $0.93\times 10^{-6}$ \\
SDSS J2222+2745 \citep{2013ApJ...773..146D}&$15\farcs1$ &2.82&0.49 & $0.67\times 10^{-6}$\\
SDSS J0909+4449 \citep{2018MNRAS.481L.136S}&$13\farcs84$&2.78&0.9&$0.84\times 10^{-6}$\\
SDSS J1326+4806 \citep{2019MNRAS.489.4741S}&$21\farcs06$&2.08&0.39&$0.74\times 10^{-6}$\\
GraL J165105.3−041725 \citep{2021ApJ...921...42S}&$10\farcs1$&1.45&0.59&$0.41\times 10^{-6}$\\
COOL J0542-2125 \citep{2023ApJ...946...63M}& $25\farcs9$&1.84&0.61&$1.09\times 10^{-6}$\\
COOL J0335-1927 \citep{2023ApJ...954L..38N}&$23\farcs3$&3.27&0.41&$1.10\times 10^{-6}$\\
\enddata
\tablecomments{The cosmological parameters for calculating the angular diameter distance are set to $(\Omega_m,\Omega_\Lambda)=(0.3,0.7)$. Assuming a transverse velocity of 500 $\mathrm{km\ s^{-1}}$ for the clusters.}
\end{deluxetable*}
The tabulated maxima span $|\Delta z_{\rm ML}|\sim(0.41\text{--}1.10)\times10^{-6}$, including $0.67\times10^{-6}$ for SDSS J1004+4112.
Figure~\ref{fig:tran_vel} shows the corresponding relation between image-to-image redshift differences and transverse velocity for SDSS J1004+4112.
Different image pairs sample different projections of the transverse motion; the attainable measurement precision is assessed separately in the following sections.

\section{Differential Redshift Measurement and Simulation}
\label{sec:Frequency Shift Measurement and Simulation}
The signal scale in Section~\ref{sec:Derivation of Moving Lens Effect} motivates a differential-redshift precision $\sigma_z$ of order $10^{-7}$--$10^{-6}$.
We estimate the idealized statistical sensitivity using mock spectra with photon noise only.
\subsection{Differential-Redshift Estimator}
We use an estimator that combines information across the spectral line profile.
This approach was developed for radial-velocity measurements in searches for extrasolar planets \citep{2001A&A...374..733B} and applied to forecasts of cosmological redshift drift \citep{2008MNRAS.386.1192L}.

Let $S_1(\lambda)$ and $S_2(\lambda)$ denote the spectra of two images with the same intrinsic line profile and a small relative wavelength shift.
Retaining the estimator's velocity-equivalent notation, the flux difference in resolution element $i$ is written to first order as
\begin{equation}
    S_{1i} - S_{2i} = \frac{\mathrm{d}S_i}{\mathrm{d}\lambda}\ \frac{\Delta v_i}{c}\ \lambda_i ,\label{eq:velocity_shift}
\end{equation}
where $\lambda_i$ is the observed wavelength and $\mathrm{d}S_i/\mathrm{d}\lambda$ is the local spectral slope.
Each resolution element supplies an estimate $\Delta v_i$, and their inverse-variance-weighted mean gives $\Delta v$.
For reporting the simulations, we use the velocity-equivalent convention $\Delta v=c\,\Delta z$ and $\sigma_v=c\,\sigma_z$; the reported shift and precision are therefore $\Delta z$ and $\sigma_z$.

Following \citet{2008MNRAS.386.1192L}, we define the weight as $\omega_i=1/\sigma^2_{v_i}$.
Propagation of the flux and spectral-slope uncertainties in Equation~\eqref{eq:velocity_shift} gives
\begin{equation}
    \sigma^{2}_{v_i} = \left[\frac{c}{\lambda_i\frac{\mathrm{d}S_i}{\mathrm{d}\lambda}} \right]^2 \ \left[\sigma^2_{1i} + \sigma^2_{2i} + \frac{(S_{1i} - S_{2i})^2}{\left( \frac{\mathrm{d}S_i}{\mathrm{d}\lambda}\right)^2}\ \sigma^2_{S^\prime_i}\right] ,\label{eq:sigma_vi2}
\end{equation}
where $\sigma_{1i}$ and $\sigma_{2i}$ are the flux uncertainties in the two spectra and $\sigma_{S^\prime_i}$ is the uncertainty in the spectral slope.
Noisy resolution elements and those with small gradients receive lower weights, allowing the combined line information to constrain shifts smaller than a resolution element without prescribing an analytic line profile for the estimator.
The first-order treatment assumes that the spectra differ only through the small wavelength shift and photon noise; changes in the intrinsic line profile are outside this forecast.
When evaluating the estimator, we exclude unstable resolution elements at the boundaries of the finite emission-line windows and use only the interior wavelength regions.
\subsection{Simulated spectrum}
In this subsection, we generate simulated high resolution spectra of the $\mathrm{C \ \Romannum{4}}$ and $\mathrm{C \ \Romannum{3}]}$ emission lines using spectral data from SDSS J1004+4112.

\begin{figure}[htbp]
    \centering
    \includegraphics[width=0.45\textwidth]{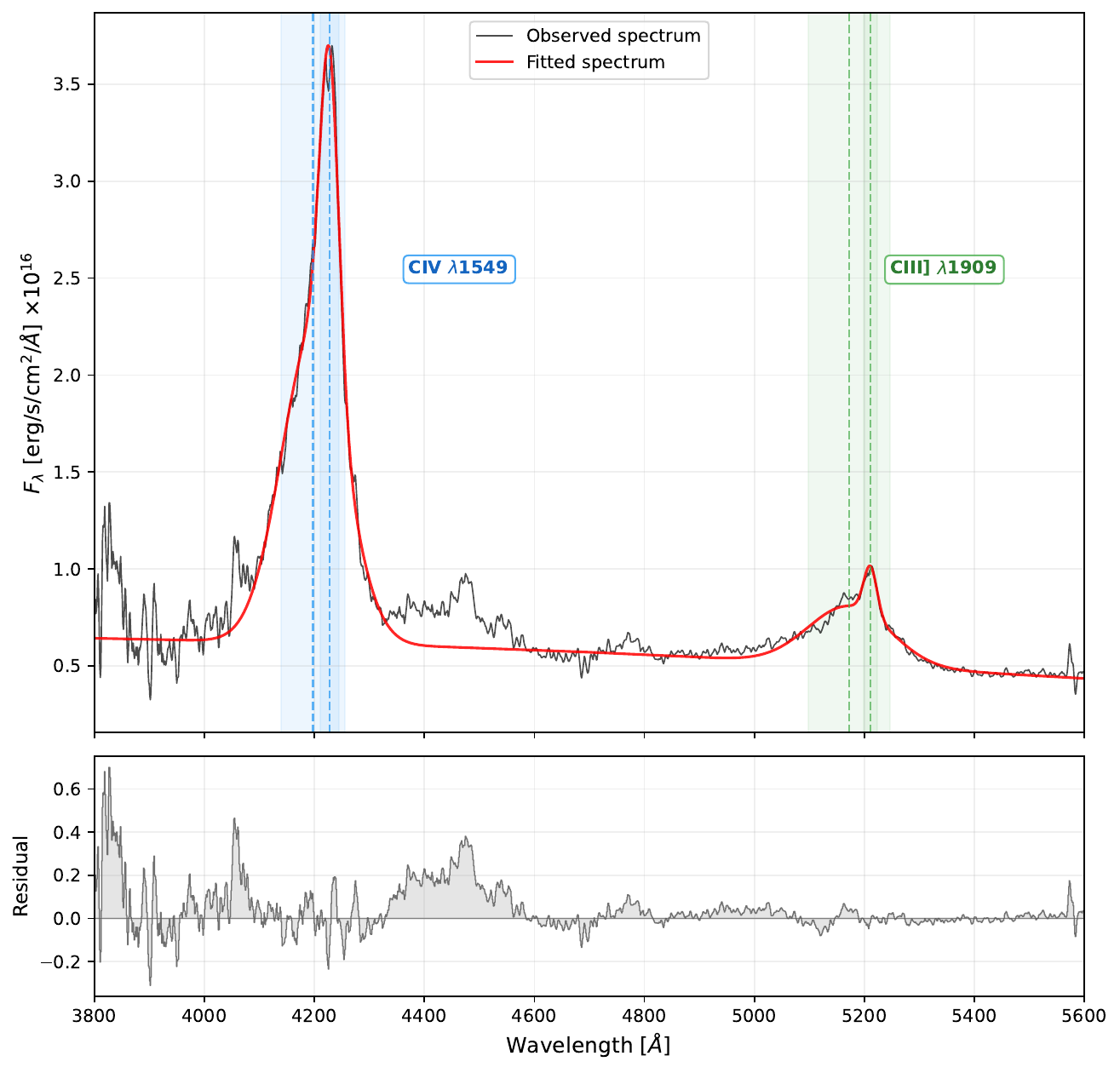}
    \caption{Illustration of the spectral fitting for SDSS J1004+4112. In the upper panel, the black line represents the observed spectrum of image A, and the red line represents the fitted spectrum. The blue and green regions represent the double Gaussian fitting results for the $\mathrm{C \ \Romannum{4}}$ and $\mathrm{C \ \Romannum{3}]}$ emission lines, respectively. The lower panel shows the residuals of the fit.}
    \label{fig:spec_fit}
\end{figure}

The method for measuring velocity drift described above not impose requirements on the type of spectrum used. However, since the weights for the differences resolution elements depend only on the measurement errors of the spectral flux and slope, any discrepancies between the profiles of two spectra will result in erroneous measurements at the given resolution element. But this will not cause large changes in the weights. In the multi-image systems under consideration, the physical separation between the light paths of different images is much larger than the characteristic size of molecular clouds or other interstellar medium component. This often results in the presence of absorption features at different wavelengths in different images. To avoid errors caused by these absorption lines, we propose to use the emission line spectra of quasars to measure the velocity drift. These emission lines have the same rest wavelength and higher flux than the continuous spectrum. We utilize spectral data of SDSS J1004+4112 B image between 3800-5500 \AA \ from \citet{2020A&A...634A..27P}, and selected the $\mathrm{C \ \Romannum{4}}$ and $\mathrm{C \ \Romannum{3}]}$ emission lines for our analysis.
We fit the continuum using a cubic spline function and remove it. The fitting windows are following spectral range: around 3900\AA , 4600\AA , and 5400\AA . Subsequently, the $\mathrm{C \ \Romannum{4}}$ and $\mathrm{C \ \Romannum{3}]}$ lines are normalized by line peak and fitted using a double gaussian profile. The fitting results are shown in Figure~\ref{fig:spec_fit}.
To simulate the redshift difference induced by moving lens, a redshift $\Delta z =1 \times10^{-6}$ is applied to generate a second set of spectra. We select the regions with a normalized intensity greater than 0.2 for the subsequent analysis.

To generate simulated spectra for a 30 meter class telescope, the following parameters in this section are determined by E-ELT and its onboard CODEX spectrograph\footnote{\url{https://www.eso.org/observing/etc/doc/elt/etc_spec_model.pdf}}.

The spectrum is then pixelized, we use four pixels as one resolution element and set a resolution of $R\ = \ 100,000$, which is the maximum of E-ELT. The $\mathrm{C \ \Romannum{4}}$ line uses the B band reference wavelength, while the $\mathrm{C \ \Romannum{3}]}$ line uses the V band reference wavelength. Defining a resolution element as the integration Wavelength range corresponding to resolution, so the range of pixel size is determined. Poisson noise is then added to the pixel. It should be noted that, since the read-out noise is smaller than photon noise, only photon noise is taken into account in the following analysis. The S/N under the photon-noise-limited is given by equation
\begin{equation}
    \frac{S}{N} = \left[ F_010^{-0.4m_{obj}}\Delta\lambda \, A \, \epsilon \,t_{int}/E_{\gamma}\right]^{\frac{1}{2}},\label{eq:SN}
\end{equation}
where $F_0$ and $m_{obj}$ are photometric zeropoint and apparent magnitude of the source in corresponding band, $\epsilon$ is total efficiency of  telescope, which is set to 0.25 for E-ELT, A is photon collecting area of the telescope, $t_{int}$ is total integration time, $E_\gamma$ is photon energy at reference wavelength. The $\Delta \lambda$ refer to the wavelength range. Since the noise is added to each pixel, the range here is taken as the size of a single pixel. It should be note that the magnitude used to calculate S/N in Equation \eqref{eq:SN} includes the continuum. After removing the continuum, this portion of noise remains in the spectrum, and its effect on the S/N must be taken into account when adding noise. 

\section{Results}
\label{sec:Results}

Table~\ref{tab:redshift_of_quasars} sets the signal scale for direct differential spectroscopic measurements in wide-separation lensed quasars.
For the adopted transverse velocity of $500\,\mathrm{km\,s^{-1}}$ and the orientation that maximizes the deflection-angle projection, the maximum moving-lens differential redshifts span $|\Delta z_{\rm ML}|\sim10^{-6}$.
These estimates motivate a target differential-redshift precision of order $10^{-7}$--$10^{-6}$.

\begin{figure}
    \centering
    \includegraphics[width=\linewidth]{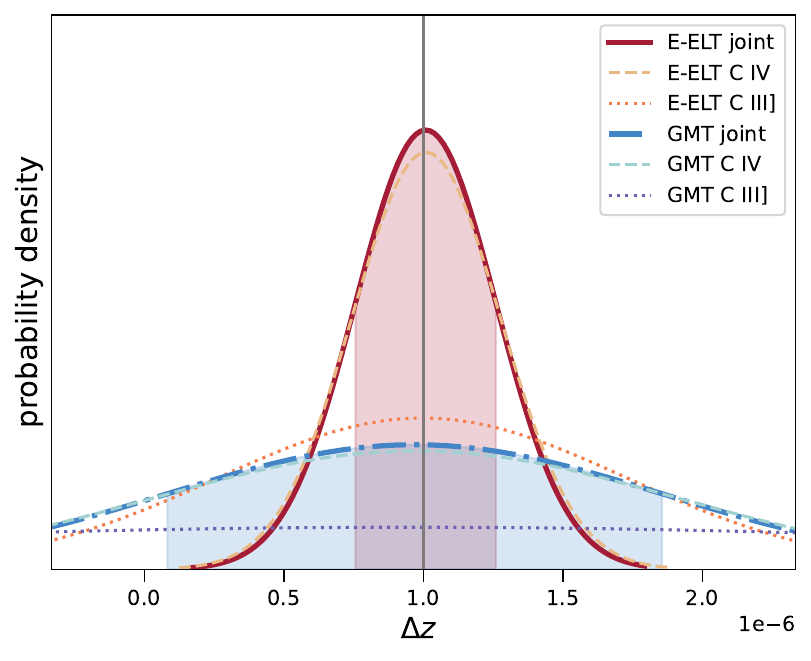}
    \caption{Photon-noise-limited Monte Carlo distributions of $\Delta z$ for SDSS J1004+4112 at $t_{\rm int}=40\,\mathrm{hr}$.
    The curves are Gaussian fits to 2000 realizations for each configuration.
    Dashed and dotted curves show $\mathrm{C \ \Romannum{4}}$ and $\mathrm{C \ \Romannum{3}]}$, respectively; the thick red solid and blue dash-dotted curves show the ELT-like and GMT joint constraints.
    The ELT-like configuration is labeled E-ELT in the legend.
    Shading marks the $1\sigma$ intervals of the joint distributions, and the gray vertical line marks $\Delta z=10^{-6}$.
    As an auxiliary comparison, the Giant Magellan Telescope (GMT) simulation, with one pixel per resolution element.
    Its weaker precision primarily reflects the smaller effective collecting aperture and photon-collecting area of the adopted GMT setup, which yield lower S/N under the same benchmark conditions.}
    \label{fig:result}
\end{figure}

We first consider the $\mathrm{C \ \Romannum{4}}$ and $\mathrm{C \ \Romannum{3}]}$ emission lines of SDSS J1004+4112 in the ELT-like configuration described in Section~\ref{sec:Frequency Shift Measurement and Simulation}.
The magnitudes of images A and B set the S/N of the two mock spectra, and the spectrum with higher S/N is used to estimate the spectral-slope uncertainty in Equation~\eqref{eq:sigma_vi2}.
For $t_{\rm int}=40\,\mathrm{hr}$ and an injected differential redshift of $\Delta z=1\times10^{-6}$, 2000 Monte Carlo realizations with photon noise only give the precisions
\[
    \begin{aligned}
    \sigma_z({\rm C\,IV})&=2.65\times10^{-7},\\
    \sigma_z({\rm C\,III]})&=7.30\times10^{-7},\\
    \sigma_z({\rm joint})&=2.51\times10^{-7}.
    \end{aligned}
\]
Figure~\ref{fig:result} shows the corresponding Monte Carlo distributions.
It should be stressed that the curves shown in figure represent the expected accuracy for a single measurement at the given frequency shift, rather than the combined constraints from 2000 simulation.

$\mathrm{C \ \Romannum{4}}$ provides a tighter constraint than $\mathrm{C \ \Romannum{3}]}$ in this simulation, consistent with its higher line intensity and effective S/N in the adopted spectra.
The joint constraint improves modestly on $\mathrm{C \ \Romannum{4}}$ alone, indicating that $\mathrm{C \ \Romannum{4}}$ supplies most of the measurement information and $\mathrm{C \ \Romannum{3}]}$ contributes an additional constraint.
Which align with the higher intensity of $\mathrm{C \ \Romannum{4}}$ compared to $\mathrm{C \ \Romannum{3}]}$, as shown in Figure~\ref{fig:spec_fit}, the peak flux of $\mathrm{C \ \Romannum{3}]}$ is only about one-third that of $\mathrm{C \ \Romannum{4}}$. This lower flux results in a substantially lower S/N for the $\mathrm{C \ \Romannum{3}]}$ emission line than for $\mathrm{C \ \Romannum{4}}$.

To examine the dependence of the measurement precision on integration time, we repeat the photon-noise-only simulations for the ELT-like configuration at $t_{\rm int}=1$, 2, 5, 10, 20, 40, 80, and 120 hr. For each integration time, 2000 Monte Carlo realizations are performed while keeping the source, telescope, and spectral-line parameters fixed.

\begin{figure}
    \centering
    \includegraphics[width=0.90\columnwidth]{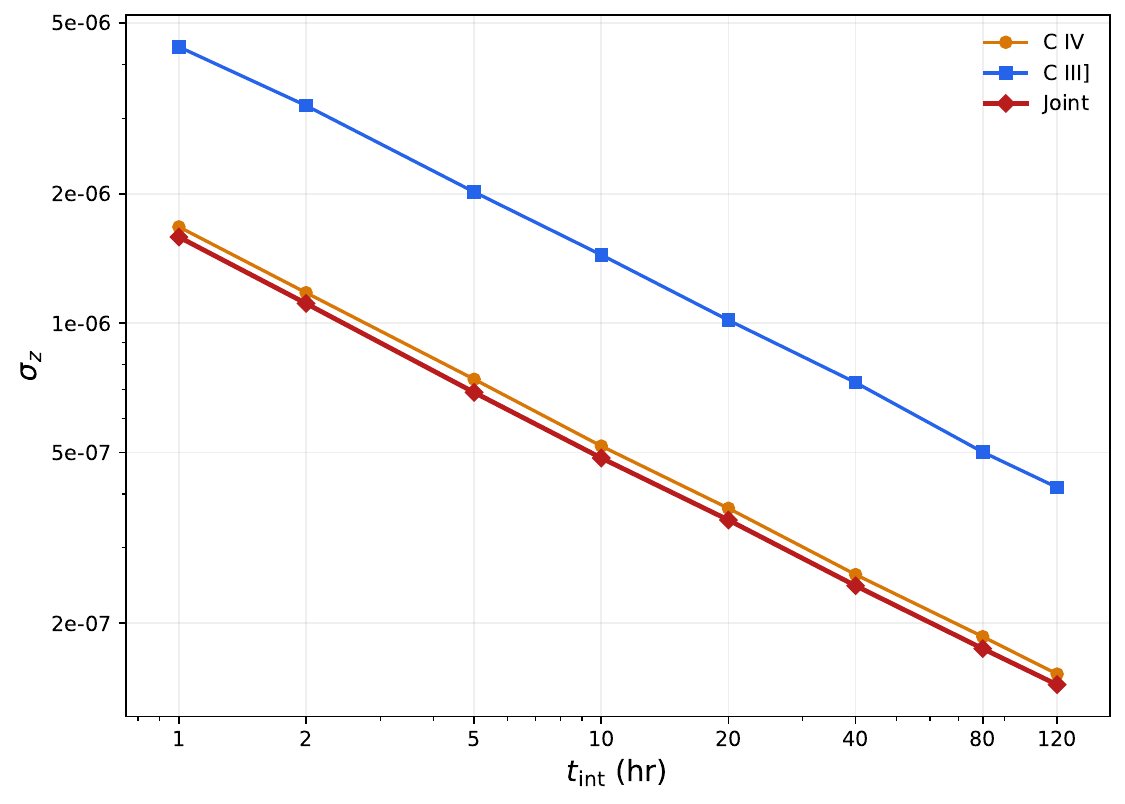}
    \caption{Photon-noise-limited Monte Carlo forecast for SDSS J1004+4112 in the ELT-like configuration.
    The horizontal axis is integration time $t_{\rm int}$ in hours and the vertical axis is differential-redshift precision $\sigma_z$, evaluated as the scatter of 2000 realizations at each integration time.
    Orange circles, blue squares, and red diamonds connected by solid lines denote $\mathrm{C \ \Romannum{4}}$, $\mathrm{C \ \Romannum{3}]}$, and the joint constraint, respectively; the joint curve is drawn with a thicker line.
    The injected differential redshift is $\Delta z=10^{-6}$ throughout.}
    \label{fig:exposure}
\end{figure}

Figure~\ref{fig:exposure} shows the resulting differential-redshift precision $\sigma_z(t_{\rm int})$, estimated from the scatter of the Monte Carlo realizations. The constraints improve continuously with increasing integration time and exhibit an approximately power-law dependence on $t_{\rm int}$, as expected from the noise model adopted in this work. Because only photon noise is included, the uncertainties associated with the individual spectra and with the spectral-slope measurements in Equation~\ref{eq:sigma_vi2} arise solely from photon noise. Meanwhile, Equation~\ref{eq:SN} shows that the signal-to-noise ratio follows a power-law dependence on the integration time $t_{\rm int}$. Therefore, within the adopted noise model, collecting more photons leads directly to improved statistical precision.

For integration times of 10 hr and longer, the joint measurement precision in the ELT-like configuration reaches the $10^{-7}$ level, entering the characteristic signal range of the moving-lens effect estimated above for wide-separation lensed quasar systems. These results suggest that direct differential spectroscopic measurements of wide-separation lensed quasars merit further investigation. However, the feasibility assessment presented here is restricted to the adopted photon-noise-limited assumptions. A more realistic evaluation will need to account additionally for instrumental calibration uncertainties and astrophysical spectral variability.

\section{Summary and Discussion}
\label{sec:Summary and Discussions}

We have evaluated wide-separation lensed quasars as targets for direct differential spectroscopic measurements of the moving-lens effect. For SDSS J1004+4112, the adopted transverse velocity of $500,\mathrm{km,s^{-1}}$ and the orientation that maximizes the signal give a maximum differential redshift of $|\Delta z_{\rm ML}|\sim6.7\times10^{-7}$.

For this system, the ELT-like photon-noise-only simulation gives a joint differential-redshift precision of $\sigma_z\sim2.5\times10^{-7}$ for $t_{\rm int}=40,\mathrm{hr}$. The $\mathrm{C \ \Romannum{4}}$ line dominates the constraint, while $\mathrm{C \ \Romannum{3}]}$ provides only a modest additional improvement.

Over the simulated range of 1--120 hr, the statistical precision improves with increasing integration time and reaches the $10^{-7}$ regime on timescales of tens of hours. Within the adopted photon-noise-only model, photon statistics alone therefore do not obviously preclude measurements at the expected signal scale.

These results describe an idealized statistical sensitivity, whereas real observations will require control of both astrophysical and instrumental systematics. One important limitation is the matching of the source epochs sampled by different lensed images. Because the images have different lensing time delays, spectra obtained simultaneously do not correspond to the same intrinsic quasar epoch. For SDSS J1004+4112, for example, the delay between images A and B has been measured to be $40.6\pm1.8$ days \citep{2008ApJ...676..761F}. Even when observations are scheduled according to the measured delay, its uncertainty and the finite observing cadence can leave a residual mismatch in source epoch. This is important because broad emission lines respond to continuum variations with finite and velocity-dependent delays: different parts of the broad-line region, and therefore different velocity components across a line profile, need not vary simultaneously \citep{2013ApJ...764...47G}. Consequently, changes in the red and blue wings or in the line asymmetry between two imperfectly matched epochs could be partially interpreted by the estimator as a wavelength shift.

Microlensing provides a distinct limitation because it does not disappear even if the spectra are matched to exactly the same source epoch. Different line-of-sight velocities can originate preferentially from different spatial regions of the broad-line region, allowing stellar microlensing to magnify different portions of an emission-line profile by different amounts. Such distortions cannot in general be removed by applying a single multiplicative flux correction. This effect is directly relevant to SDSS J1004+4112: asymmetric enhancement of the blue wings of high-ionization lines was identified in early spectroscopic monitoring \citep{2004ApJ...610..679R}, and subsequent studies have confirmed long-lived, velocity-dependent microlensing distortions of the C~IV profile, with the blue and red wings experiencing different magnifications \citep{2023A&A...672A..45H,2024A&A...682A..57F}. These profile changes can therefore introduce additional scatter or bias into a differential-redshift measurement based on flux differences across the line.

Instrumental effects must also be controlled in an observational implementation. Spectra obtained at different epochs should be transformed consistently to the Solar-system barycentric frame; barycentric corrections are standard in precision spectroscopy and can be calculated to substantially better precision than required here when the observing time and target coordinates are accurately known \citep{2014PASP..126..838W}. Residual wavelength-calibration drifts and changes in the instrumental line-spread function may nevertheless be important when comparing spectra taken at different epochs. Detector contributions such as read noise and dark current, neglected in the present photon-noise-only simulations, should also be included in a complete exposure-time calculation, although they are expected to be less important in regimes where the source photon counts dominate.

This work therefore establishes a photon-noise-limited statistical benchmark for direct image-to-image spectroscopy of the moving-lens effect and shows that photon noise alone can reach the relevant signal scale. The observational feasibility will depend on controlling quasar variability, microlensing-induced line-profile distortions, and spectroscopic calibration systematics. If these effects can be modeled, monitored, or controlled to the required level, image-to-image spectroscopy would provide a transverse-motion probe of individual strong-lens systems complementary to statistical CMB/LSS measurements \citep{2026arXiv260518938H}.

\paragraph{Note added.}
After completing this manuscript, we became aware of independent work
proposing closely related ideas \citep{2026arXiv260938457W}.

\begin{acknowledgments}
This work was supported by the National Key Research and Development Program of China Grant Nos. 2023YFC2206702, and 2021YFC2203001; National Natural Science Foundation of China under Grants Nos. 12405050, 1243300112322301, and 12275021.

The authors acknowledge the use of ChatGPT during the revision of this manuscript solely for language editing. No AI tool was used to generate scientific content or produce figures.
\end{acknowledgments}

\bibliography{sample701}{}
\bibliographystyle{aasjournalv7}

\end{document}